\RequirePackage[2020-02-02]{latexrelease}
\documentclass[pre,aps,twocolumn,showpacs]{revtex4}
\usepackage{epsfig}
\include{graphics}

\begin{document}

\title{Distribution of currents in a system of active particles with biased hops and Ising interactions}

\author{Evgeniy Khain}
\email{khain@oakland.edu}
\affiliation{Department of Physics, Oakland University, Rochester, MI 48309, USA}

\author{Shriram Ramesh}
\email{shriramu@umich.edu}
\affiliation{College of Engineering, University of Michigan, Ann Arbor, MI 48109, USA}

\author{Vladimir Chernyak}
\email{chernyak@chem.wayne.edu}
\affiliation{Department of Chemistry, Wayne State University, Detroit, MI 48202, USA}

\begin{abstract}
We consider an ensemble of particles performing a biased random walk on a two-dimensional lattice. Due to the adhesion between particles (nearest-neighbor Ising interactions), the probability of hopping depends on the number of neighbors and the strength of adhesive interaction. The asymmetry in the hopping directions leads to a macroscopic current of particles through the system. We theoretically compute the distribution of currents and verify the results in stochastic particle simulations. To make theoretical progress, one has to determine the probabilities of various particle configurations such as single particles, pairs, and different configurations of three-particle clusters. These probabilities are theoretically derived by formulating rate equations for the concentrations of various structures resembling Becker-Döring cluster equations. One intriguing finding is that the concentration of left-handed triplets is not equal to the concentration of right-handed triplets for non-zero asymmetry of hops and non-zero interaction between particles.
\end{abstract}

\maketitle

\section{Introduction}

An ensemble of random walkers that hop on a lattice is a common model for the migration of motile cells on a substrate \cite{Simpson}. Assuming (nearest-neighbor) Ising interactions between the random walkers mimics the adhesion between neighboring cells, and the corresponding model has been employed to describe collective cell migration \cite{migration} and cell clustering \cite{clustering}. A crucial modification of the model consists in introducing a directional bias in particle migration, specifically, a biased random walk. The model was first introduced in the 1980s in a different context of stochastic lattice gas under the influence of a uniform external field \cite{KLS1,KLS2,KLS3}.

The model without a bias can be formulated also in the language of spins with nearest-neighbor Ising interactions: an occupied site corresponds to a spin ``up", an empty site corresponds to a spin ``down", and the adhesion parameter (see below) is related to the ratio of magnetic coupling $J$ to the thermal energy $k_B T$. As the result, for high cell adhesion parameter (above the critical threshold), cells form clusters, similarly to formation of magnetic domains below the critical temperature. It turns out that the bias affects the phase transition threshold and the forming clusters are elongated in the direction of the bias \cite{KLS1,KLS2,KLS3}; multiple-strip configurations and coarsening below the critical temperature have been investigated \cite{ordering}. The asymmetry in hopping leads to a steady current of particles throughout the system \cite{KLS1,KLS2,KLS3,flux}. In stochastic simulations, one obtains different overall currents in every run for the same values of parameters. This work aims at finding the distribution of currents as a function of parameters for low adhesion (or high temperature in the language of spins), far away from the ordering phase transition.

In the steady state, every configuration of particles in the system occurs with a certain probability. It turns out that the asymmetry of the hopping changes the probabilities of various configurations. To determine these probabilities, we construct rate equations for the concentrations of such structures motivated by the Becker-Döring cluster equations used to describe coarsening and Ostwald ripening \cite{Becker}. Then we map this multi-particle dynamics into a single effective random walker and employ the Fokker–Planck equation to theoretically compute the distribution of currents. Finally, we compare the theoretical findings with the results of our stochastic particle simulations.

The rest of the paper is organized as follows. In Section II, we present the model for a biased random walk with Ising interactions on a two-dimensional lattice. Section III describes the main theoretical results and compares these results with stochastic particle simulations. Section IV includes a brief discussion and summary of our results.

\section{The Model}

Consider a two-dimensional square lattice $L$ by $L$ where each lattice site can be empty or occupied by a single particle. Periodic boundary conditions are implemented in both $x$ and $y$ directions: if a particle hops out of the system, it re-enters from the other side. At each step, a randomly selected particle might hop to a neighboring empty site or remain at rest. Due to the adhesion between neighboring particles, the probability of hopping away decreases with a greater number of neighbors and a stronger adhesion. 

In addition, the particle hopping is biased: the chance of hopping to the right is larger than the chance of hopping left. As a result, a macroscopic current of particles flows throughout the system. The described model has three parameters: $\rho$, the average dimensionless concentration of particles (the total number of particles divided by the total number of sites), $0 \leq q \leq 1$, the adhesion strength between particles, and $0 \leq \mu \leq 1$, the degree of asymmetry in hopping. These parameters define the probabilities of hopping upward $p_{up}$, downward $p_{down}$, to the right $p_{right}$, and to the left $p_{left}$:

\begin{eqnarray}
p_{up} = \frac{1}{4}\,(1-q)^n, \\ \nonumber
p_{down} = \frac{1}{4}\,(1-q)^n, \\ \nonumber
p_{right} = \frac{1+\mu}{4}\,(1-q)^n, \\ \nonumber
p_{left} = \frac{1-\mu}{4}\,(1-q)^n,
\end{eqnarray}
where $n$ is the number of nearest neighbors. $q=0$ means no adhesive interactions (so the number of neighbors does not play a role in this case) and $\mu=0$ corresponds to the case of no bias. 

When performing stochastic particle simulations, the time is advanced after each single particle step by $1/k$, where $k=\rho\,L^2$ is the number of particles in the system. In every simulation, we kept track of the number of hops to the right and to the left and computed the overall current $I$ defined to be just the total number of right hops minus the total number of left hops during the simulation time $t_f$. For each set of parameters ($\rho$, $q$, $\mu$), we performed between $5000$ to $50000$ runs, computing the current $I$ and the speed $v=I/t_f$ in every run.

Below we describe the main results of stochastic particle simulations. Figures $1$ and $2$ show the distribution of speeds for zero asymmetry ($\mu=0$, Figure $1$) and nonzero asymmetry ($\mu=0.1$ and $\mu=0.3$, Figure $2$): symbols show the results of stochastic particle simulations, while solid and dashed curves show the theoretical results (derived below). The inset of Figure $1$ presents the width of the distribution for $\mu=0$ as a function of $q$: the blue curve shows the theoretical results, while the red squares correspond to stochastic particle simulations. The inset of Figure $2$ presents the scaled (by $\mu$) average speed as a function of $\mu$ for different values of $q$: the curves show the theoretical results, while the symbols correspond to stochastic particle simulations. One can see that some deviations from the theoretical results occur only at larger values of $q$.

\begin{figure}[ht]
    \centering
    \includegraphics[width=3.6 in]{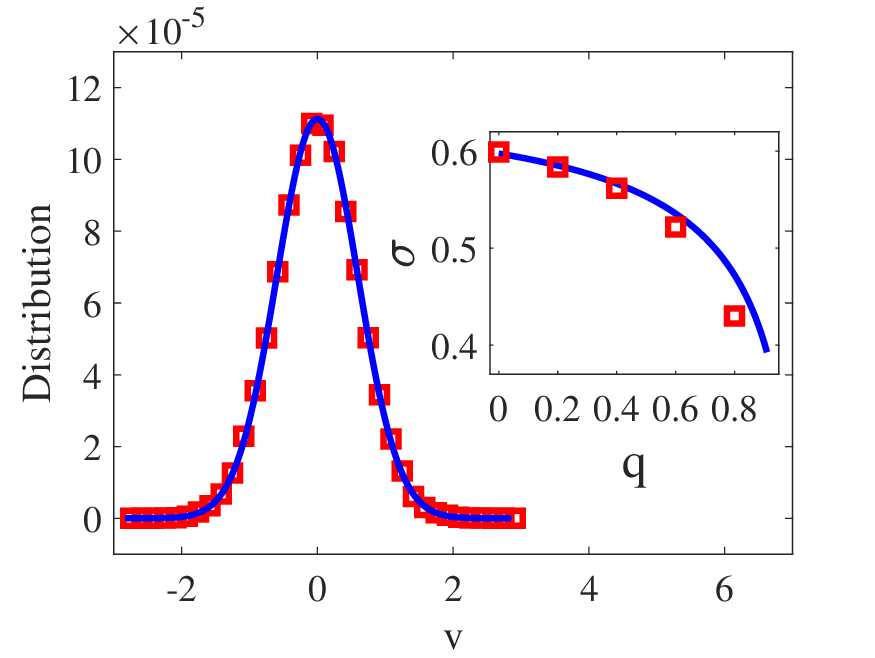}
    \caption{Distribution of speeds for symmetric hops. Red squares show the results of stochastic particle simulations, while the blue solid line corresponds to the theoretical calculations. The parameters are $n=0.05$, $q=0$, and $\mu=0$. The inset shows the width of the distribution as a function of the adhesion parameter $q$ from both stochastic simulations (red squares) and theory (blue solid line). The parameters are $\rho=0.05$ and $\mu=0$.}
\end{figure}

\begin{figure}[ht]
    \centering
    \includegraphics[width=3.6 in]{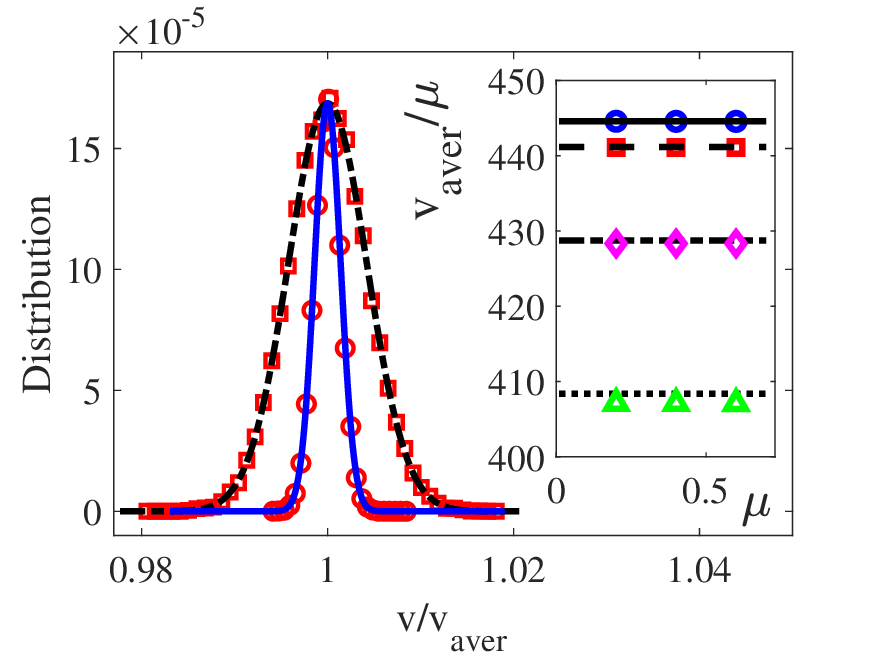}
    \caption{Distribution of speeds (rescaled by the average speed) for asymmetric hops. A more narrow distribution shown by the blue solid line (theory) and red circles (stochastic particle simulations) corresponds to a higher asymmetry of hops ($\mu=0.3$), while a wider distribution shown by the black dashed line (theory) and red squares (stochastic particle simulations) corresponds to a smaller asymmetry of hops ($\mu=0.1$). Other parameters are $\rho=0.005$ and $q=0.05$. The inset shows the rescaled average speed as a function of the asymmetry parameter $\mu$ for different values of the adhesion parameter $q$. The curves show the theoretical results, while the symbols represent the results of stochastic particle simulations. The values of the adhesion $q$ are (from top to bottom): $q=0.05$, $q=0.2$, $q=0.5$, and $q=0.7$.}
\end{figure}

\begin{figure}[ht]
    \centering
    \includegraphics[width=2.3 in]{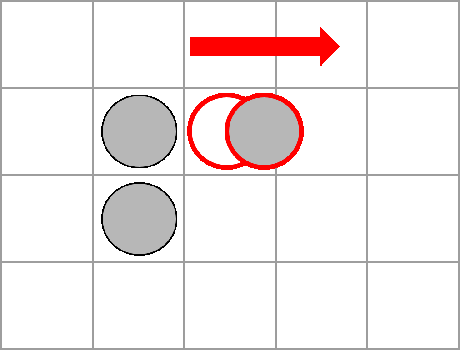}
    \includegraphics[width=2.3 in]{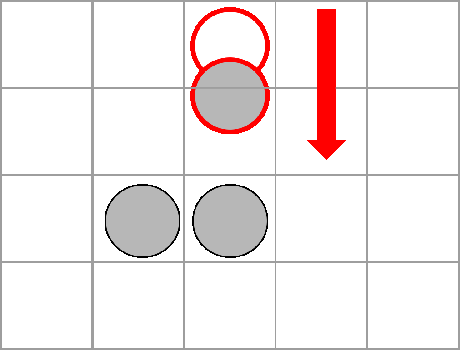}
    \caption{Upper panel: A schematic representation of the detachment process: a single particle detaches from the right-handed triplet leaving a pair and a single. Lower panel: A schematic representation of the attachment process: a single particle attaches to a pair forming a left-handed triplet.}
\end{figure}

In the next section we consider dilute systems and derive a theory that reproduces the distribution of currents in the system and its dependence on the governing parameters. Since different particle configurations contribute differently to the overall current, affecting the chances of hopping to the right and to the left, our first goal is to determine the occurrence probability of these configurations.

\section{Densities of Clusters of Various Sizes and Configurations}

Consider a low-density regime where clusters of $4$ particles and larger occur very rarely and can be ignored. What is the chance that a chosen particle is a single particle, belongs to a pair, or belongs to a triplet (a cluster of $3$ particles)? Since different triplets contribute differently to the current, one also needs to distinguish between straight, left-handed, and right-handed triplets, Figure $3$. In the following, we write down equations for the density of single particles, $\rho_1$, the density of pairs, $\rho_2$, and the densities of left-handed, $\rho_{3lh}$, right-handed, $\rho_{3rh}$, and straight, $\rho_{3s}$, triplets. These equations include transitions between various clusters that occur when a hopping particle detaches from or attaches to a certain cluster (configuration). Figure $3$ shows examples of such attachment and detachment processes.

The probability of attachment of a single particle to a pair (Figure $3$, the lower panel) is proportional to the density of singles, $\rho_1$, and pairs, $\rho_2$, respectively. The figure shows a downward jump (occurring with a probability of $1/4$) that creates a left-handed triplet. Three other jumps of a single particle towards the pair create a left-handed triplet: one upward hop (a probability of $1/4$) and two left hops (a probability of $(1-\mu)/4$ each). However, the pair can be positioned not only horizontally but also vertically. In this case, a left-handed triplet can be created as a result of one downward jump (a probability of $1/4$), one upward jump (a probability of $1/4$) and two right jumps (a probability of $(1+\mu)/4$ each). Assuming that the concentration of vertical pairs equals the concentration of horizontal pairs, we get the following term on the right-hand side of the equation for $d\rho_{3lh}/dt$: $\rho_1\,\rho_2$. The same term should enter the equations for $d\rho_{1}/dt$ and $d\rho_{2}/dt$ with the minus sign. Clearly, not all of these attachments of a single particle to a pair result in the formation of a left-handed trippet: in the same way we computed the probabilities of the creation of right-handed and straight triplets.

Now we consider the detachment process, in which the right-hand triplet disappears, and a single particle and a pair appear in the system. First, the detachment from the right-handed triplet (Figure $3$, the upper panel) is proportional to its concentration, $\rho_{3rh}$. Next, in the diagram, the rightmost particle moves right, which happens with a probability of $(1-q)(1+\mu)/4$, as it detaches from one neighbor. This rightmost particle can also jump upward, which happens with a probability of $(1/4)(1-q)$. In addition, the downward particle might jump either downward (a probability of $(1/4)(1-q)$) or to the left (a probability of $(1-q)(1-\mu)/4$). The right-handed triplet can have its rightmost particle in the lower line; the calculation of the detachment probabilities is done in a similar way. Overall, we get the following term on the right-hand side of the equation for $d\rho_{3rh}/dt$: $-\rho_{3rh}(1-q)$. Again, the same term should enter the equations for $d\rho_{1}/dt$ and $d\rho_{2}/dt$ with the plus sign.

Notice that while considering detachment, we did not check for whether or not the hopping single particle has neighbors after the detachment. Multiplying by the density correction, $(1-\rho)^3$ would ensure that the $3$ spots around the single will be empty. However, in the limit of low density, $0<\rho<<1$, the density of pairs is proportional to $\rho_1^2$, and the density of triplets scales as $\rho_1^3$. Since terms of order $\rho_1^4$ and above are neglected and the detachment term is already proportional to $\rho_{3rh}$, the density correction should be ignored.

Considering all possible attachments and detachments (not shown) that change the concentrations of singles, pairs, and various triplets, we arrive at the following rate equations:

\begin{widetext}
\begin{eqnarray}
%rho_1
\frac{d\rho_1}{dt} = \frac{3}{4}\rho_{3lh}{(1-q)}^{2}(2+\mu) + \frac{3}{4}\rho_{3rh}{(1-q)}^{2}(2-\mu) + \frac{3}{2}\rho_{3s} {(1-q)}^{2} - 9{\rho_1}^3 \\ \nonumber + \rho_{3lh} (1-q) + \rho_{3rh} (1-q) + \frac{3}{2} \rho_{3s}(1-q)\rho_2(1-q)(3-7\rho) - \frac{7}{2}\rho_1\rho_2 - 6{\rho_1}^2(1-2\rho), \\ \nonumber
\frac{d\rho_2}{dt} = \rho_{3lh}(1-q) + \rho_{3rh}(1-q) + \frac{3}{2}\rho_{3s}(1-q) - \frac{1}{2}\rho_2(1-q)(3-7\rho) - \frac{7}{2}\rho_1\rho_2 + 3{\rho_1}^2(1-2\rho), \\ \nonumber
\frac{d\rho_{3lh}}{dt} = \rho_{3rh}(1-q)\frac{1+\mu}{4} + \rho_{3lh}(1-q)\frac{1-\mu}{4} - \rho_{3lh}(1-q)^2\frac{2+\mu}{4} + {\rho_1}^3\frac{2-\mu}{2} - \rho_{3lh}(1-q) + \rho_1\rho_2, \\ \nonumber
\frac{d\rho_{3rh}}{dt} = \rho_{3lh}(1-q)\frac{1-\mu}{4} - \rho_{3rh}(1-q)\frac{1+\mu}{4} - \rho_{3rh}(1-q)^2\frac{2-\mu}{4}
+ {\rho_1}^3\frac{2+\mu}{2} - \rho_{3rh}(1-q) + \rho_1\rho_2, \\ \nonumber
\frac{d\rho_{3s}}{dt} = - \frac{1}{2}\rho_{3s}(1-q)^2 - \frac{3}{2}\rho_{3s}(1-q) + {\rho_1}^3 + \frac{3}{2}\rho_1\rho_2.
\end{eqnarray}
\end{widetext}

Assuming a steady state, we can find the concentrations of various clusters as a function of parameters $\rho$, $q$, and $\mu$. For unbiased hops, $\mu=0$, we obtain

\begin{eqnarray}
\frac{\rho_2}{\rho_1^2} &=& \frac{2}{1-q}, \\ \nonumber
\frac{\rho_{3s}}{\rho_1^3} &=& \frac{2}{(1-q)^2}, \\ \nonumber
\frac{\rho_{3rh}}{\rho_1^3} &=& \frac{2}{(1-q)^2}, \\ \nonumber
\frac{\rho_{3lh}}{\rho_1^3} &=& \frac{2}{(1-q)^2},
\end{eqnarray}

It is clear from these results that our procedure is valid when the parameter $\epsilon=\rho/(1-q)$ is much smaller than $1$. Indeed, the ratio $\rho_2/\rho_1$ is proportional to $\epsilon$; the ratio of the concentration of triplets (of any kind) and the concentration of singles is proportional to $\epsilon^2$, while the terms of order $\epsilon^3$ and higher (clusters of $4$ particles and larger) are neglected.

For a nonzero asymmetry of hops, $\rho_1$, $\rho_2$, and $\rho_{3s}$ remain the same, but $\rho_{3lh}$ and $\rho_{3rh}$ change. The difference in concentrations of these triplets, $\Delta\rho = \rho_{3lh}-\rho_{3rh}$ depends both on $\mu$ and $q$

\begin{equation}
\frac{\Delta\rho}{\rho_1^3}=\frac{\mu\,q}{(1-q)^2}\frac{16(3-q)}{4(4-q)(3-q)+\mu^2(1-q^2)}.
\end{equation}

\begin{figure}[ht]
    \centering
    \includegraphics[width=3.5 in]{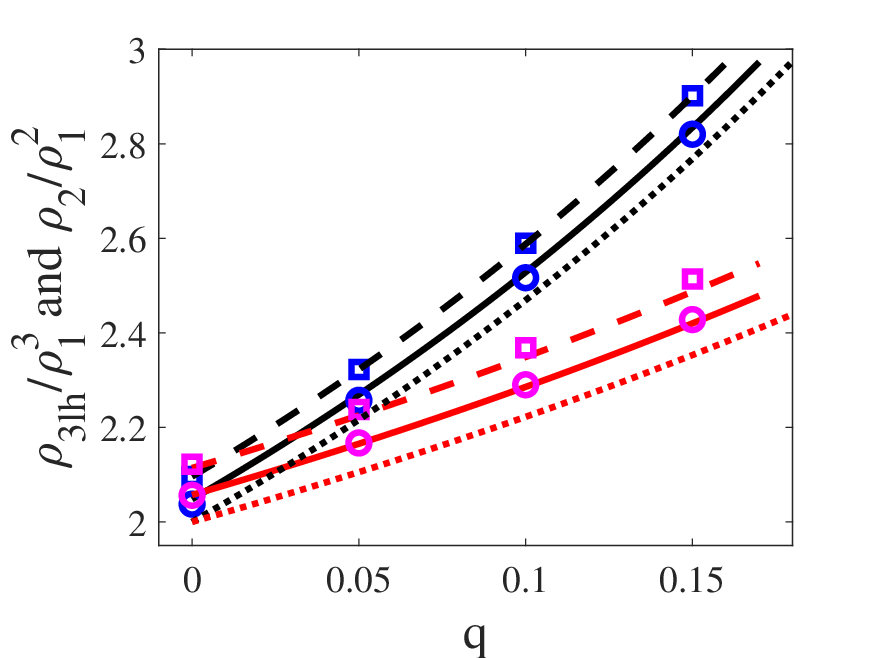}
    \includegraphics[width=3.5 in]{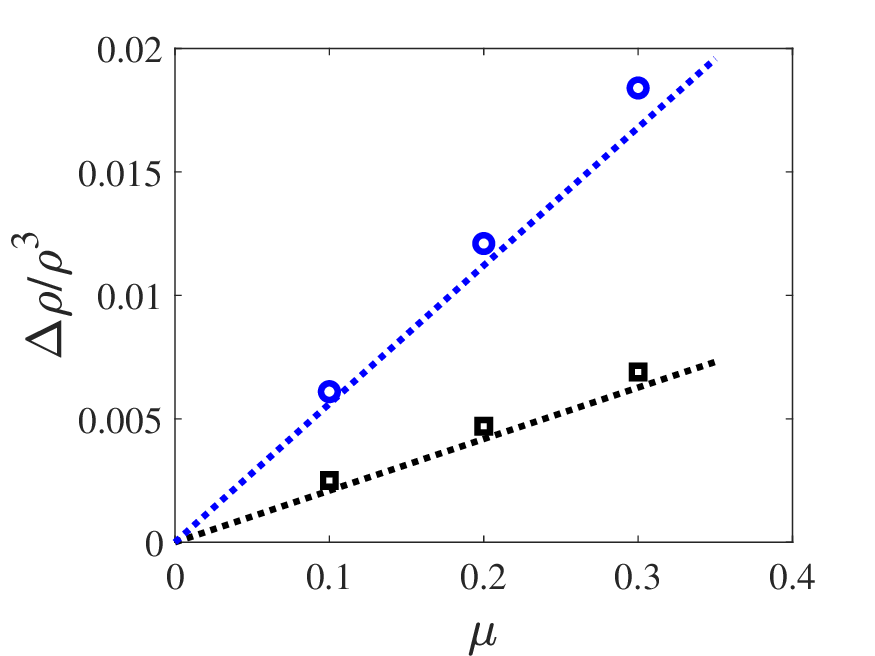}
    \caption{Upper panel: The scaled densities of the left-handed triplets (steeper curves) and pairs (lower curves) for zero bias as a function of $q$. The dashed lines correspond for the theoretical results in the dilute density limit, the solid lines (theory) and the circles (stochastic particle simulations) correspond for $\rho=0.005$, while the dashed lines (theory) and the squares (stochastic particle simulations) correspond for $\rho=0.01$, see text for more detail. Lower panel: The scaled difference between the concentrations of the left-handed and right-handed triplets ($\Delta\rho = \rho_{3lh}-\rho_{3rh} $) as a function of $\mu$ for two values of $q$. The parameters are: $\rho=0.005$ and $q=0.05$ correspond to the blue dotted line (theory) and blue circles (stochastic simulations), $\rho=0.01$ and $q=0.02$ correspond to the black dotted line (theory) and black squares (stochastic simulations)}
\end{figure}

Importantly, the change in the concentration of these triplets requires not only biased hops (nonzero $\mu$), but also Ising interactions (nonzero $q$). Figure $4$ compares these theoretical expressions with the results of stochastic particle simulations. The lower panel shows the value of the scaled $\Delta\rho$ as a function of $\mu$ for different values of the overall density $\rho$ and the adhesion parameter $q$, a good agreement between the theoretical results (dotted curves) and the results of stochastic particle simulations (symbols) can be observed. The upper panel shows the ratio $\rho_{3lh}/\rho_1^3$ (the upper black curves and the corresponding blue symbols) and the ratio $\rho_{2h}/\rho_1^2$ (the lower red curves and the corresponding magenta symbols) as a function of the adhesion parameter $q$. In both cases, the dotted curves show the theoretical expression, while the solid curves and dashed curves correspond to the same expression with a density correction: $$\frac{\rho_2}{\rho_1^2} = \frac{2}{1-q}(1+5.7\,\rho)$$ and $$\frac{\rho_{3lh}}{\rho_1^3} = \frac{2}{(1-q)^2}(1+4.8\,\rho).$$ The solid lines (theory) and circles (stochastic particle simulations) are plotted for $\rho=0.005$, while the dashed lines (theory) and squares (stochastic particle simulations) are plotted for $\rho=0.01$. The figure indicates that our results are valid in the limit of low densities, where the corrections can be ignored.

Once the concentrations of these structures are known, we can compute the current. Once a particle is chosen at random (which happens with probability $\rho$), what are the probabilities of hopping to the right and to the left? This depends on which structure the particle belongs to: whether or not it is a single particle, whether or not it belongs to a pair, left-handed, right-handed, or straight triplets. Once this is known, we find the chance for a particle in each of these structures to hop to the right. For example, the contribution of single particles to $p_{right}$ would be $\frac{1+\mu}{4}\rho_{1}$. However, the contribution of pairs is slightly more involved: only one of two particles can hop to the right in the horizontal pair, but both particles in a vertical pair can hop to the right. Taking into account singles, pairs and three possible triplets, the overall (scaled) probabilities of hopping to the right and to the left become
%\begin{widetext}
\begin{eqnarray}
\frac{4\,\rho}{1+\mu}p_{r} &=& [(1-q)^2+(1-q)]\rho_{3lh} \\ \nonumber
&+& 2(1-q)\rho_{3rh} + [1.5(1-q)+0.5(1-q)^2]\rho_{3s} \\ \nonumber
&+& 1.5(1-q)\rho_{2}+\rho_1, \\ \nonumber
\frac{4\,\rho}{1-\mu}p_{l} &=& 2(1-q)\rho_{3lh} + [(1-q)^2 + (1-q)]\rho_{3rh} \\ \nonumber
&+& [1.5(1-q)+0.5(1-q)^2]\rho_{3s} \\ \nonumber
&+& 1.5(1-q)\rho_{2}+\rho_1.
\end{eqnarray}
%\end{widetext}

Above, we defined the current $I$ to be the total number of hops to the right minus the total number of hops to the left. Every time one of the particles in the system jumps to the right, $I$ increases by one, and every time any particle in the system jumps to the left, $I$ decreases by one. Therefore, the problem of finding the current becomes analogous to a single effective random walker (representing the $\rho L^2$ spatially distributed particles) hopping to the right with probability $p_{r}$ and hopping to the left with probability $p_{l}$. The corresponding distribution of currents $p(I, t)$ will be described by the Fokker–Planck equation with a certain diffusion coefficient $D$ and a certain drift coefficient $v_d$.

\begin{equation}
\frac{\partial p}{\partial t}=D\frac{\partial^2 p}{\partial I^2} - v_d\frac{\partial p}{\partial I}.
\end{equation}

Finding the diffusion coefficient and the drift velocity is a standard mathematical task. Consider random variables $I_1$, $I_2$, ... $I_N$ such that $I_i$ describes the hopping of a particle on time step $i$ such that $I_i=1$ with probability $p_{r}$, $I_i=-1$ with probability $p_{l}$, and $I_i=0$ with probability $1 - p_{r} - p_{l}$. The overall current during these $N$ time steps is $I = I_1+I_2+...+I_N$. The average is $<I>=N\,<I_i> = N(p_{r} - p_{l})$. Now, $<I^2>$ = $N\,<(I_i)^2> + N(N-1)<I_i\,I_j>$ = $N(p_{r} + p_{l}) + N(N-1)(p_{r} - p_{l})^2$. Therefore, $<I^2> - <I>^2$ = $N(p_{r} + p_{l} - (p_{r} - p_{l})^2)$. As the result, $D = \rho L^2 (p_{r} + p_{l} - (p_{r} - p_{l})^2)/2$ and $v = \rho L^2 (p_{r}-p_{l})$.

The solution for this equation is well-known
\begin{equation}
p(I,t)=\frac{1}{\sqrt{4\pi\,D\,t_f}}\, e^{-(I-v_d\,t_f)^2/(4\,D\,t_f)}
\end{equation}

Now we can refer again to Figures $1$ and $2$ presented at the beginning of the paper. These figures show the resulting distribution of speeds, where the speed is the current divided by the total time, $v=I/t_f$. An excellent agreement between the theoretical results (curves) and the results of stochastic particle simulations (symbols) can be observed both for a zero asymmetry of hops ($\mu=0$, Figure $1$) and for a nonzero $\mu$ ($\mu=0.1$ and $\mu=0.3$, Figure $2$).

\section{Summary and discussion}

In this work, we considered a model for particles with Ising interactions performing a biased random walk on a lattice: the probability to hop to the right is different from the probability to hop to the left. This model should be distinguished from the active Ising model \cite{active}, where a spatial site can have any number of particles (spins), particles with positive and negative spins preferentially hop in the opposite directions, and Ising interactions occur only among particles on the same spatial site.

We have shown that the asymmetry in hopping ($\mu\neq0$) in the presence of Ising interactions between particles ($q\neq0$) changes the probabilities of various configurations of particles; this is a novel effect not present in one-dimensional systems \cite{Chernyak}. In particular, the concentration of right-handed triplets increases and the concentration of left-handed triplets decreases, compared to the case of no bias. In other words, for $q=0$ (no Ising interactions), the instantaneous snapshot of the system does not show the viewer if particles preferentially hop right or left. For nonzero interactions, however, the relative abundance of these two structures (as seen on a snapshot) implies that particles preferentially hop in a certain direction.

To determine the probabilities of different configurations, we formulated rate equations for the concentrations of various particle structures, similar to the Becker-Döring cluster equations. Our equations are more involved as we accounted not only for the number of particles in a cluster, but also for various particle configurations having the same number of particles. To construct a closed system of equations, we considered the case of low density $\rho$ and low adhesion $q$, neglecting the existence of clusters of size $4$ and larger. We checked that this is a very reasonable approximation as long as the parameter $\rho/(1-q)$ is much smaller than $1$.

Once the concentrations of singles, pairs, and various configurations of triplets were computed, the distribution of currents (speeds) was derived from the effective Fokker–Planck equation. It is quite remarkable that there is a way to find the entire distribution of currents in a complex many-particle system. The resulting theoretical description is in excellent agreement with the results of stochastic particle simulations (Figures $1$ and $2$). However, the tails of the distribution (the chance of having a particularly large speeds) cannot be described by the Fokker–Planck equation. The correct description requires employing special techniques borrowed from the theory of rare events (see Ref. \cite{rare} for review). Such methods were used to find rare large clusters in interacting lattice gas systems \cite{rare1}, rare transitions in driven granular media \cite{rare2}, population extinction \cite{Dykman}, rare large front speeds \cite{rare3}, and rare large topological currents \cite{rare4}; finding rare large currents in our system is an interesting avenue of future research.

Another possible extension of this work is (initially) organizing different currents in different parts of the system, this might lead to spatially inhomogeneous steady states. This can be done, for example, by having different temperatures in different spatial locations \cite{twofluxes}.

There can be many situations where living cells have a preferential direction of migration on a substrate on top of the random component. One such phenomenon is chemotaxis \cite{chemotaxis}, in which cells preferentially migrate toward (or away from) a high concentration of certain chemicals. It would be interesting to experimentally detect the difference in the concentrations of various small cell clusters with and without chemotaxis. However, it is not clear whether or not it is possible to organize periodic boundary conditions (and steady states) in such experimental systems.

\end{document}